\documentclass[aps,prb,twocolumn,showpacs]{revtex4}
\usepackage{graphicx}
\usepackage{dcolumn}
\usepackage{bm}

\begin{document}
\title{Coupled electron and phonon transport in one-dimensional atomic junctions}
\author{J. T. L\"u}
\email{tower.lu@gmail.com}
\author{Jian-Sheng Wang}
\homepage[]{http://staff.science.nus.edu.sg/~phywjs/}
\affiliation{
Center for Computational Science and Engineering and Department of Physics,
National University of Singapore, Singapore 117542, Republic of Singapore
}
\date{\today}
\begin{abstract}
Employing the nonequilibrium Green's function method, we develop a fully
quantum mechanical model to study the coupled electron-phonon transport in
one-dimensional atomic junctions in the presence of a weak electron-phonon
interaction. This model enables us to study the electronic and phononic
transport on an equal footing. We derive the electrical and energy currents of
the coupled electron-phonon system and the energy exchange between them. As
an application, we study the heat dissipation in current carrying atomic
junctions within the self-consistent Born approximation, which guarantees
energy current conservation. We find that the inclusion of phonon transport
is important in determining the heat dissipation and temperature change of
the atomic junctions.
\end{abstract}
\maketitle

\section{Introduction}
\label{sec:intro}

The electronic transport and phononic transport in meso- and nano-structures have 
attracted a great deal of interest in the past two decades, although their
development is not so parallel sometimes. These structures display important
quantum effects due to the confinement in one or more
directions\cite{ciraci-review}.  The quantized electrical
conductance\cite{PhysRevLett.60.848} was observed much earlier than that of
the thermal conductance\cite{PhysRevLett.81.232} mainly due to the
difficulty in measuring the thermal transport properties. Electrons
and phonons are not two isolated systems. Their interactions are important
for both electronic and phononic transport. With the development of both
fields there arises the requirement to study the coupled electron-phonon
transport from time to time. When studying electronic transport problems,
one usually assumes that electrons interact with some phonon bath where the
phonons are in their thermal equilibrium state characterized by the Bose
distribution.  This simple assumption is not able to give satisfactory
results in some cases where the phonons are driven out of equilibrium
by the electrons. This is especially true in places where the
thermal conductance is low or the phonon relaxation is
slow\cite{pop:155505,lazzeri:236802}.  To take into account the
nonequilibrium phonon effect, one usually introduces into the electronic
transport formalism some phenomenological parameters that describe the
phonon relaxation process. In engineering applications, as the size of the
electronic devices decreases to nanoscale, the heat dissipation and
conduction in these structures become critical issues, which may influence
the electronic properties dramatically\cite{cahill:793}. Only studying the
electronic transport is not enough in these cases.  On the other hand, heat transport in
one-dimensional (1D) structures has received considerable attention
recently\cite{cahill:793,lepri-review,li:015121}.  Fourier's law of heat
conduction is no longer valid in many 1D systems. The
microscopic origins of the macroscopic Fourier's law remain one of the most
frustrating problems in nonequilibrium statistical mechanics. Since the electrons and phonons both
contribute to the heat conduction, their relative roles in many
nanostructures are still not clear. Especially in
semiconductors, which one carries the majority of the thermal current is not a
trivial problem. To answer these questions, we need some general models,
which take into account the electron, phonon transport, and their mutual
interactions.

Theoretically, although the development of electronic transport in 1D
structures has been very striking, that of the phononic transport is
relatively slow. Classical molecular dynamics (MD) and the Boltzmann-Peierls
equation are the widely used methods in phononic transport. MD method is not
accurate below the Debye temperature, while the Boltzmann-Peierls equation can
not be used in nanostructures without translational invariance. In both cases,
the quantum effect becomes important\cite{ciraci-review}.  Only recently,
the nonequilibrium Green's function
method\cite{keldysh01,kadanoff1,jauho-book,datta1}, which has been widely
used to study the electronic transport, has been applied to study the quantum phononic
transport\cite{PhysRevB.63.125415,PhysRevB.68.245406,yamamoto:255503,dhar:085119,wang:033408}.
As far as we know, the study of the coupled electronic and phononic transport in
nanostructures is
rare\cite{MGalperion07-mtj,ryndyk:045420,auer:165409,lazzeri:165419}.  In
Ref.~\onlinecite{ryndyk:045420}, the authors considered the nonequilibrium phonons in
molecular transport junctions. Galperin and co-authors analyzed the heat
generation and conduction in molecular systems\cite{MGalperion07-mtj}.  In
this paper, using the nonequilibrium Green's function method, we study the
coupled electronic and phononic transport in 1D atomic junctions.  The
formalism is similar to that of Ref.~\onlinecite{MGalperion07-mtj}. In our model
the electron subsystem is described by a single-orbital tight-binding
Hamiltonian, and the phonon subsystem is described in a harmonic
approximation. We assume that the electron-phonon interaction is weak so
that the perturbative treatment is valid. The strong-interaction
case is the scope of future work. 

The rest of the paper is organized as follows. In Sec.~\ref{sec:model}, we
introduce the 1D model system, and derive expressions for the electrical,
energy current of the coupled electron-phonon system.  In Sec.~\ref{sec:app}
we show the heat generation in one- and two-atom structures under different
model parameters. Sec.~\ref{sec:con} is the conclusion. In Appendix
\ref{sec:se}-\ref{sec:ana} we give some technical details of our derivation.

\section{Coupled electronic and phononic transport}
\label{sec:model}
\subsection{The Hamitonian}
\label{ham}
\begin{figure}[!htbp]
\includegraphics[scale=0.55]{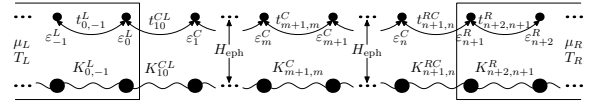}
\caption{\label{fig:show}Shematic diagram of the 1D coupled electron-phonon system and the parameters used in the model. The big dots in the bottom line represent atoms, while the small dots in the upper line represent electron states. They are coupled via the electron-phonon interaction.}
\end{figure}

Our model system is an infinite 1D atomic chain as shown in Fig.~\ref{fig:show}.  The electrons
and atoms are only allowed to move in the longitudinal direction.  We treat
the atoms as coupled harmonic oscillators, and take into account their
nearest neighbour interactions up to the second order. We assume that there
is only one single electronic state for each atom and take into account hopping 
transitions between the nearest states. This corresponds to a single-orbital
tight-binding model. Also, we assume that there is only one spin state for each orbital. Following Caroli\cite{caroli1}, we divide
the whole system into one central region and two semi-infinite leads, which
act as electrical and thermal baths (Fig.~\ref{fig:show}). The Hamiltonian of the whole system is
\begin{eqnarray}
H&=&\sum_{\alpha=L,C,R;\beta={\rm e,ph}}H^{\alpha}_{\beta} \nonumber \\
 &+&\sum_{\alpha=L,R;\beta={\rm e,ph}}\left(H_\beta^{\alpha C} + H_\beta^{C\alpha}\right) + H_{{\rm eph}}.
\label{eq:ham0}
\end{eqnarray}
The electron-phonon interaction Hamiltonian $H_{{\rm eph}}$ is non-zero only in
the central region. The electron Hamiltonian reads
\begin{equation}
	H^\alpha_{\rm e} = \sum_{i}\varepsilon^\alpha_i c^{\dagger\alpha}_ic^{\alpha}_i + \sum_{|i-j|=1}t^\alpha_{ij}c^{\dagger\alpha}_ic^{\alpha}_j,
	\label{eq:eham0}
\end{equation}
where $c^{\dagger\alpha}_{i}$ and $c^{\alpha}_i$ are the electron creation
and annihilation operators. $\varepsilon^\alpha_i$ is the electron onsite
energy, and $t^\alpha_{ij}$ is the hopping energy between adjacent states.
$i$ and $j$ run over the sites in the $\alpha$ region. The
coupling Hamiltonian with the leads is
\begin{equation}
	H_{\rm e}^{LC} = \sum_{ij} {t_{ij}^{LC}} c^{\dagger L}_ic^C_j,
	\label{eq:ell0}
\end{equation}
and
\begin{equation}
	H_{\rm e}^{CR} = \sum_{ij} t_{ij}^{CR} c^{\dagger C}_ic^R_j.
	\label{eq:elr0}
\end{equation}
$H_{\rm e}^{CL}$ and $H_{\rm e}^{RC}$ have similar expressions. We also have $t^{\alpha
C} = {t^{C\alpha}}^{\dagger},~\alpha = L, R$. For our 1D tight-binding
model, $t^{\alpha C}$ has only one non-zero element. If we label the central atoms with indices $1$ to $n$ as shown in Fig.~\ref{fig:show}, the non-zero elements will be $t^{LC}_{01}$, $t^{CL}_{10}$, $t^{RC}_{n+1,n}$, and $t^{CR}_{n,n+1}$.

The phonon Hamiltonian is 
\begin{equation}
	H^\alpha_{\rm ph} = \frac{1}{2}\sum_{i} \dot{u}^\alpha_i\dot{u}^\alpha_i + \frac{1}{2}\sum_{|i-j|=0,1} u^\alpha_i K^\alpha_{ij} u^\alpha_j.
	\label{eq:phham0}
\end{equation}
$u^\alpha_{i}$ and $\dot{u}^\alpha_i$ are the mass-renormalized atom
displacement and momentum operator. $K^\alpha_{ii} = 2K^\alpha_0/m_i^\alpha$,
and $K^\alpha_{ij} = -K^\alpha_0/\sqrt{m_i^\alpha m_j^\alpha}~(i\ne j)$. Here $K^\alpha_0$ is the spring constant, and $m_i^\alpha$ is the mass of the $i$th atom in the $\alpha$ region. Like the electrons, the coupling Hamiltonian with the leads is
\begin{equation}
	H_{\rm ph}^{LC} = \frac{1}{2}\sum_{ij} u^{L}_iK^{LC}_{ij}u^C_j,
	\label{eq:phll0}
\end{equation}
and
\begin{equation}
	H_{\rm ph}^{CR} = \frac{1}{2}\sum_{ij} u^{C}_iK^{CR}_{ij}u^R_j.
	\label{eq:phlr0}
\end{equation}
We also have $K^{C\alpha} = {K^{\alpha C}}^\dagger$. The non-zero elements
are $K^{LC}_{01}$, $K^{CL}_{10}$, $K^{RC}_{n+1,n}$, and $K^{CR}_{n,n+1}$.

The electron-phonon interaction is included within the adiabatic
Born-Oppenheimer approximation. First, the electron subsystem is solved with
all the atoms in their equilibrium positions. Then, the isolated phonon
subsystem is considered. After that, the electron-phonon interaction is turned
on by allowing the atoms to oscillate around their equilibrium positions.
Within this picture, the electron-phonon interaction is\cite{jauho-book}
\begin{equation}
	H_{\rm eph} = \sum_{i,j,k} M_{ij}^k c^{\dagger}_ic_ju_k.
	\label{eq:ephham0}
\end{equation}
The interaction matrix element is $M_{ij}^k = \left\langle
i\left|\frac{\partial H_{\rm e}}{\partial u_k}\right|j\right\rangle$. All the
operators in Eq.~(\ref{eq:ephham0}) are in the central region, so we omitted
the superscript $C$.  In our model, the electron operators are in the second
quantization, while that of the phonons are in the first quantization.

\subsection{Green's functions}
\label{subsec:green}
The nonequilibrium Green's function method for the electronic transport is
discussed in Refs.~\onlinecite{keldysh01,kadanoff1,jauho-book,datta1}, and that for the phononic transport
in Refs.~\onlinecite{PhysRevB.63.125415,PhysRevB.68.245406,yamamoto:255503,dhar:085119,wang:033408}. Here we
concentrate on the electron-phonon interactions. The definition of the
electron contour-ordered Green's function is $G_{jk}(\tau,\tau ')=-i\langle \mathcal{T}\{ c_j(\tau) c_k^\dagger(\tau ')\}\rangle$, and the phonon counterpart is 
$D_{jk}(\tau,\tau ')=-i\langle \mathcal{T}\{ u_j(\tau) u_k(\tau ')\}\rangle$. Here
$\tau$ is time on the Keldysh contour, and $\mathcal{T}\{\cdots\}$ is the contour-ordered operator. We set $\hbar=1$ throughout the formulas. Without the electron-phonon interaction, the isolated
electron and phonon problem can be solved exactly. We denote these 
Green's functions as $G_0(\tau,\tau ')$ and 
$D_0(\tau,\tau ')$, respectively.  In our case, it is convenient to write the
Hamiltonians as matrices and work in the energy space.  The electron retarded and advanced 
Green's functions are
$G^r_0(\varepsilon) = {G^a_0}^\dagger(\varepsilon) = \left[(\varepsilon + i\eta)I - H^C_{\rm e} - \Sigma^{r}_L(\varepsilon)- {\Sigma}^{r}_R(\varepsilon)\right]^{-1}$.
$I$ is an identity matrix, and $\eta\rightarrow 0^+$. The retarded self-energy $\Sigma^{r}_\alpha=
t^{C\alpha}g^{r}_\alpha t^{\alpha C}$ is due to the interactions
with the lead $\alpha$. The retarded Green's function of the semi-infinite lead $g^r_\alpha$ can be obtained analytically (Appendix \ref{sec:se}). The ``less than'' Green's function is given by $G^<_0 = G^r_0(\Sigma^{<}_L+\Sigma^{<}_R)G^a_0$, where $\Sigma^{<}_\alpha = -f^{\rm e}_\alpha(\Sigma^{r}_\alpha-\Sigma^{a}_\alpha)$. $f_\alpha^{\rm e}$ is the Fermi-Dirac distribution. The 
phonon retarded and advanced Green's functions are\cite{note2}
$D^r_0(\omega) = {D^a_0}^\dagger(\omega) = \left[(\omega+i\eta)^2I-K^C-\Pi^r_L(\omega)-\Pi^r_R(\omega)\right]^{-1}$.
The lead retarded self-energy is
$\Pi^r_\alpha(\omega)=K^{C\alpha}d^{r}_\alpha(\omega)K^{\alpha C}$.
$d_\alpha^r$ also has analytical expression (Appendix \ref{sec:se}).
The phonon ``less than'' Green's function is $D^<_0 =
D^r_0(\Pi_L^<+\Pi_R^<)D_0^a$, where $\Pi_\alpha^< =
f_\alpha^{\rm ph}(\Pi_\alpha^r-\Pi_\alpha^a)$. $f^{\rm ph}_\alpha$ is the Bose
distribution function. 

Knowing the bare electron and phonon Green's functions $G_0$ and $D_0$, we
can include their interaction as perturbation. Following the standard
procedure of nonequilibrium Green's function method, we can express this
interaction as self-energies. The full Green's functions are obtained from
the Dyson equation, e.g., for electrons $G^{r,a} =
G_0^{r,a}+G_0^{r,a}\Sigma^{r,a}_{\rm eph}G^{r,a}$, and $G^< = G^r \Sigma^<_{\rm t}
G^a$. $\Sigma^<_{\rm t} = \Sigma^<_{\rm eph}+\Sigma^<_{\rm L}+\Sigma^<_{\rm R}$ is the total self-energy. Keeping the lowest non-zero order (the second order) of the
self-energies, we have two (Hartree- and Fock-like) terms for the electrons,
and one polarization term for the phonons. This is the so-called Born approximation (BA)\cite{jauho-book}. 
The Fock self-energies are
\begin{equation}
	\Sigma_{mn}^{F,<}(\varepsilon) =i M_{mi}^k \int {G_0^<}_{ij}(\varepsilon-\omega){D_0^<}_{kl}(\omega)\frac{d\omega}{2\pi}M_{jn}^{l},
	\label{eq:ephfocklesse}
\end{equation}
and
\begin{eqnarray}
	\Sigma_{mn}^{F,r}(\varepsilon)&=&iM_{mi}^k \int \frac{d\omega}{2\pi}\left[{G_0^r}_{ij}(\varepsilon-\omega){D_0^<}_{kl}(\omega)\right. \nonumber \\
	&&+{G_0^<}_{ij}(\varepsilon-\omega){D_0^r}_{kl}(\omega) \nonumber \\
	&&\left.+ {G_0^r}_{ij}(\varepsilon-\omega){D_0^r}_{kl}(\omega)\right]M_{jn}^{l}.
	\label{eq:ephfockretardede}
\end{eqnarray}
The ``less than'' Hartree self-energy is zero, and the retarded one is
\begin{equation}
	\Sigma_{mn}^{H,r} = -i M_{mn}^i {D_0^r}_{ij}(\omega '=0)M_{kl}^j\int {G_0^<}_{lk}(\varepsilon)\frac{d\varepsilon}{2\pi}.
	\label{eq:ephhatreefretarded}
\end{equation}
This term is a constant for all energies, which represents a static
potential due to the presence of phonons. The self-energies for the phonons are 
\begin{equation}
	\Pi_{mn}^<(\omega) = -i M_{lk}^m\int \frac{d\varepsilon}{2\pi}{G_{0}^<}_{ki}(\varepsilon){G_{0}^>}_{jl}(\varepsilon-\omega)M_{ij}^n,
	\label{eq:phselessf}
\end{equation}
and
\begin{eqnarray}
	\Pi_{mn}^r(\omega)&=&-i M_{lk}^m\int \frac{d\varepsilon}{2\pi}\left[{G_{0}^r}_{ki}(\varepsilon){G_{0}^<}_{jl}(\varepsilon-\omega)\right. \nonumber \\
	&&\left. +{G_{0}^<}_{ki}(\varepsilon){G_{0}^a}_{jl}(\varepsilon-\omega)\right]M_{ij}^n.
	\label{eq:phserf}
\end{eqnarray}
In Eqs.~(\ref{eq:ephfocklesse}-\ref{eq:phserf}), sum over internal indices
is assumed. The self-consistent Born approximation (SCBA) is obtained by
replacing all the bare Green's functions $G_0$ and $D_0$ in Eqs.
(\ref{eq:ephfocklesse}-\ref{eq:phserf}) with the full $G$ and
$D$\cite{jauho-book}. In Appendix~\ref{sec:cons}, we show that the
SCBA fulfills the electrical and energy current conservation, while BA
fails.

\subsection{The electrical and energy current}
\label{subsec:current}
The electrical and energy current can be expressed by the Green's
functions. The electrical current
out of the lead $\alpha$ is\cite{PhysRevLett.68.2512,jauho-book}
\begin{equation}
	J_\alpha = e \int \frac{d\varepsilon}{2\pi}~\mathrm{Tr}\{G^>(\varepsilon) \Sigma^<_\alpha(\varepsilon) - G^<(\varepsilon)\Sigma^>_\alpha(\varepsilon)\}.
	\label{eq:ecur}
\end{equation}
The electron energy current is
\begin{equation}
	J^{\rm E,e}_\alpha =  \int \frac{d\varepsilon}{2\pi} \varepsilon~\mathrm{Tr}\{G^>(\varepsilon) \Sigma^<_\alpha(\varepsilon) - G^<(\varepsilon)\Sigma^>_\alpha(\varepsilon)\}.
	\label{eq:eecur}
\end{equation}
The electron heat current is obtained from Eqs.
(\ref{eq:ecur}-\ref{eq:eecur}) as $J^{\rm h,e}_\alpha =
J^{\rm E,e}_\alpha-\mu_\alpha J_\alpha/e$.  $\mu_\alpha$ is the lead chemical potential. The derivation of the phonon energy
current runs parallel with that of the electrons\cite{wang:033408}
\begin{equation}
	J^{\rm E,ph}_\alpha = -\int \frac{d\omega}{4\pi} \omega~\mathrm{Tr}\{D^>(\omega) \Pi^<_\alpha(\omega) - D^<(\omega)\Pi^>_\alpha(\omega)\}.
	\label{eq:ephcur}
\end{equation}
For phonons the energy current is the same as the heat current. When there
is no electron-phonon interaction, the electron energy current is
conserved throughout the structure. So is the phonon energy current. 
In the presence of such an interaction, only the total energy
current is conserved due to the energy exchange between them. The phonons do not carry charges, so in
both cases the electrical current is conserved. Since we can't get
the exact self-energies in most cases, we need some approximations. Properly defined self-energies should fulfill the
electrical and energy current conservation
\begin{equation}
\sum_\alpha J_\alpha = 0,
	\label{eq:econ}
\end{equation}
\begin{equation}
\sum_\alpha (J_\alpha^{\rm E,e}+J_\alpha^{\rm E,ph})=0,
	\label{eq:eecon}
\end{equation}
where $\alpha$ runs over all the leads. We justify that the SCBA
fulfills these conservation laws, while the BA fails to conserve
the energy current (Appendix B). Provided we satisfy these conservation
laws, we can write the electrical and energy current in symmetric forms. 
The electrical current is
\begin{equation}
\label{eq:ecur1}
J = e\int \frac{d\varepsilon}{2\pi} \tilde{T}^{\rm e}(\varepsilon)\left[f^{\rm e}_L(\varepsilon)-f^{\rm e}_R(\varepsilon)\right].
\end{equation}
The transmission coefficient reads
\begin{eqnarray}
\label{eq:etransmission}
\tilde{T}^{\rm e}&=&\mathrm{Tr}\left\{\frac{1}{2}\left [G^r(\Gamma_L+\frac{1}{2}\Gamma_{\rm eph}-S^{\rm e})G^a\Gamma_R \right.\right.\nonumber \\
 &&\left.\left.+G^r\Gamma_LG^a(\Gamma_R+\frac{1}{2}\Gamma_{\rm eph} + S^{\rm e})\right]\right\},
\end{eqnarray}
where $S^{\rm e}$ is
\begin{equation}
\label{eq:etransmission2}
S^{\rm e} = \frac{\frac{1}{2}(f^{\rm e}_R+f^{\rm e}_L)\Gamma_{\rm eph} +
i\Sigma_{\rm eph}^<}{f^{\rm e}_L-f^{\rm e}_R}.  
\end{equation}
$\Gamma_\alpha=i(\Sigma^r_\alpha-\Sigma^a_\alpha),~\alpha=L,R$ is the
electron level-width function. $\Gamma_{\rm eph} =
i(\Sigma^r_{\rm eph}-\Sigma^a_{\rm eph})$ is due to the electron-phonon interaction.
The total energy current is
\begin{eqnarray}
\label{eq:etcur}
J^{\rm E}&=&\int \frac{d\varepsilon}{2\pi} \varepsilon \left\{\tilde{T}^{\rm e}(\varepsilon)\left[f^{\rm e}_L(\varepsilon)-f^{\rm e}_R(\varepsilon)\right]\right.\nonumber\\
&&\left.-\frac{1}{2}\tilde{T}^{\rm ph}(\varepsilon)\left[f^{\rm ph}_L(\varepsilon)-f^{\rm ph}_R(\varepsilon)\right]\right\}.
\end{eqnarray}
The phonon transmission coefficient is 
\begin{eqnarray}
\label{eq:phtransmission}
\tilde{T}^{\rm ph}&=&\mathrm{Tr}\left\{\frac{1}{2}\left [D^r(\Lambda_L+\frac{1}{2}\Lambda_{\rm eph}-S^{\rm ph})D^a\Lambda_R \right.\right. \nonumber \\
&&\left.\left. +D^r\Lambda_LD^a(\Lambda_R+\frac{1}{2}\Lambda_{\rm eph} + S^{\rm ph})\right]\right\},
\end{eqnarray}
where $S^{\rm ph}$ is
\begin{equation}
\label{eq:phtransmission2}
S^{\rm ph} = \frac{\frac{1}{2}(f^{\rm ph}_R+f^{\rm ph}_L)\Lambda_{\rm eph} - i\Pi_{\rm eph}^<}{f^{\rm ph}_L-f^{\rm ph}_R}.
\end{equation}
$\Lambda_\alpha = i(\Pi^r_\alpha-\Pi^a_\alpha)$ is the phonon
level-width function. $\Lambda_{\rm eph} = i(\Pi^r_{\rm eph}-\Pi^a_{\rm eph})$ is due to
the electron-phonon interaction. Eqs.  (\ref{eq:ecur1}-\ref{eq:phtransmission2}) are
the generalization of the Caroli formula\cite{caroli1} to include the
electron-phonon interaction.

\section{Heat generation in current carrying 1D atomic junctions}
\label{sec:app}
As an application of the formalism in Sec.~\ref{sec:model}, we study the
heat dissipation in current-carrying 1D atomic
junctions\cite{PhysRevB.46.4757,todorov98-lhi,MJMontgomery02-pdi,APHorsfield06-tto,PhysRevLett.88.216803,pecchia:035401,MGalperion07-mtj,sun-2006}.
In the presence of potential difference between the two leads, there will be
electrical current flowing between them. When the electrons
pass the central region, there is energy exchange between the electron and
phonon systems. The energy dissipated into the phonon system makes the atom
temperature higher than that of the leads if it is not
efficiently conducted to the leads.  If the electron-phonon
interaction is weak, the energy dissipated into the phonon system is only a
small fraction of the electron energy current. But this small fraction still influences the
transport properties of the atomic junction and even leads to junction
breakup\cite{PhysRevLett.86.3606,ventra:176803}, especially when
the thermal conductance is low.  Different models have been used to study
the local heating effect. Some simply assume that the phonons are in their
thermal equilibrium states\cite{sun-2006}. Some take into account the phonon
transport by using the rate equations\cite{pecchia:035401} or other
semi-classical
models\cite{todorov98-lhi,MJMontgomery02-pdi,PhysRevLett.86.3606}. Few of
them take into account the quantum effect in heat
transport\cite{ycchen03-lhi,MGalperion07-mtj}. Our model
treats the electron and phonon transport on an equal quantum-mechanical
footing, and includes their interactions self-consistently. The heat
generation is given by (Eq.~(\ref{eq:selconte2}))
\begin{eqnarray}
	\label{eq:heatgen}
	Q&=&i\int \frac{d\varepsilon}{2\pi} \int \frac{d\omega}{2\pi} \omega \nonumber \\
	&&\times \left[G^>_{nm}(\varepsilon)M_{mi}^kD_{kl}^<(\omega)G^<_{ij}(\varepsilon-\omega)M_{jn}^l\right].
\end{eqnarray}
At zero temperature, we can get an analytical expression
Eq.~(\ref{eq:heat3}) for a single-atom structure by using the bare Green's
functions $G_0$ and $D_0$ in Eq.~(\ref{eq:heatgen}) (Appendix
\ref{sec:ana}).  Equation~(\ref{eq:heat3}) can reproduce most qualitative
features of heat generation in a single atom, except that it does not take
into account heat conduction in the phonon system. 

\begin{figure}[!htbp]
\includegraphics[scale=1.2]{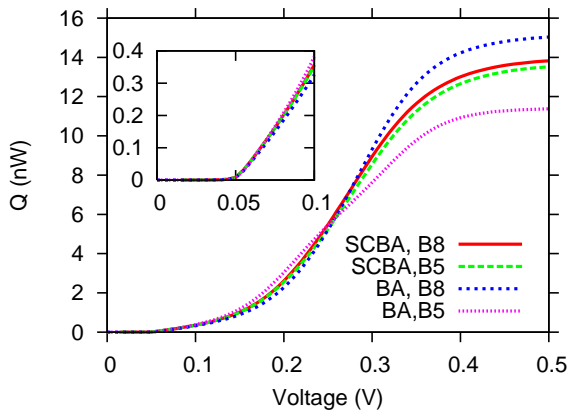}
\caption{\label{fig:scvsnosc}Comparison of different methods to compute the
heat generation in a single atom structure. The four curves correspond to
results from Eqs.~(\ref{eq:selcontfock}) and (\ref{eq:sephonon}) under BA
and SCBA, respectively.  If we label this single atom as index $1$, its
electronic onsite energy is written as $\varepsilon^C_1=0.1$ eV. The
onsite energy of the leads is $\varepsilon^L =\varepsilon^R=0$ eV. The
hopping energy is $t^L_{ij}=t^R_{ij}=0.5$ eV.  The non-zero electronic
coupling with the lead is $t^{CL}_{10}=t^{RC}_{21}=0.1$ eV.  The
matrix element of the single atom is $K^{C}_{11} = 0.654$ eV/(\AA$^2\cdot$ amu). The spring
constant between the lead atoms is $K_{ij}^L=K_{ij}^R=0.654$
eV/(\AA$^2\cdot$amu).  The non-zero atomic coupling with the leads is
$K^{CL}_{10}=K^{RC}_{21}=0.127$ eV/(\AA$^2\cdot$amu).} 
\end{figure}

We first study the case where the lead energy band is wide compared to the
voltage applied to the structure. For most metallic leads, this condition
should hold. In the weak electron-phonon coupling regime, the Born
approximation should give acceptable results for the heat generation,
although physically it is not a \emph{good} approximation.  Figure
\ref{fig:scvsnosc} shows the heat generation of a single atom ($n=1$ in
Fig.~\ref{fig:show}) computed using Eqs.  (\ref{eq:selcontfock}) and
(\ref{eq:sephonon}) under BA and SCBA, respectively. The parameters used in the calculation are stated in the figure caption. With these parameters, the electron energy band is in
the range $-1 \le \varepsilon \le 1$ eV. The chemical potential of each lead
is zero in equilibrium. The phonon energy is approximately $\omega = 0.05$
eV.  In all the results presented in this section, the temperature is $T = 4.2$
K, the electron-phonon coupling matrix $M = 0.08$
eV/(\AA$\cdot$amu$^{\frac{1}{2}})$.  The cut-off energy of the electron system
is $2.1$ eV, and the phonon system is $0.2$ eV. The energy spacing is
discretized into grids of $1$ meV.  Equation~(\ref{eq:selcontfock}) gives
the energy decrease of the electron system, while Eq.~(\ref{eq:sephonon})
gives the energy increase of the phonon system. Numerical results from
Eq.~(\ref{eq:selcontfock}) and Eq.~(\ref{eq:sephonon}) under SCBA have some
slight discrepancy. This is due to numerical inaccuracies.  But most of
the discrepancy under BA comes from the difference between the bare and the
full Green's functions, which may become even larger for some parameters. So
BA should be used with care in the study the energy exchange between the
electron and phonon system. We also note that although Eqs.
(\ref{eq:selcontfock}) and (\ref{eq:selconte}) are equivalent, numerical
result from Eq.~(\ref{eq:selcontfock}) is unstable in many cases.  The
reason is that the energy exchange between the electron and the phonon
system is only a small fraction of the total electrical energy current.
Equation~(\ref{eq:selcontfock}) is the difference between two large numbers,
so our numerical integration has to be accurate enough to get a reasonable
result\cite{pecchia:035401}.  On the contrary, Eq.~(\ref{eq:heatgen}) is
much more stable since we have got the difference analytically. All the
results presented below use this equation.

\begin{figure}[!htbp]
\includegraphics[scale=1.2]{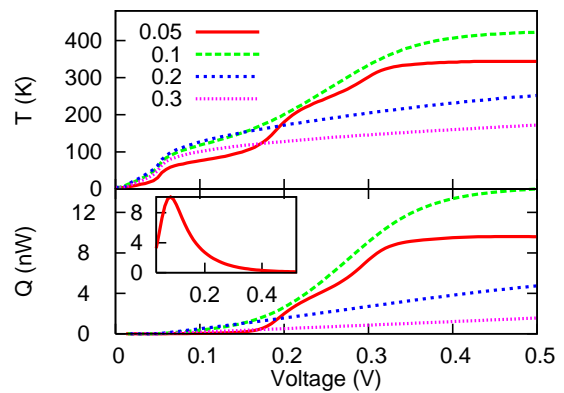}
\caption{\label{fig:ecouple}Heat generation $Q$ and the atom temperature $T$
under different electron coupling strength $t^{CL}_{10}=t^{RC}_{21}= 0.05$,
$0.1$, $0.2$, and $0.3$ eV, respectively. Other parameters are the same with
Fig.~\ref{fig:scvsnosc}. The inset shows the heat generation as a function
of electron coupling strength at an applied bias $V = 0.3$ V.}
\end{figure}

From Fig.~\ref{fig:scvsnosc}, we can see two threshold values in heat
generation. 
The first one corresponds to the onset of phonon emission.
Under low temperatures, the equilibrium phonon occupation is very small, so
the phonon absorption process seldom takes place. If the applied bias is
smaller than the phonon energy, electrons don't have enough energy to emit
one phonon. So the heat generation is zero. Once the applied voltage is
larger than the phonon energy, phonon emission turns on. The heat generation
increases almost linearly with the applied bias (inset of Fig.~\ref{fig:scvsnosc}).  This is different from the electrical current, which increases smoothly in this regime. The second threshold value corresponds to the alignment
of the left lead chemical potential with the electron onsite energy $eV =
2\varepsilon_0$ (positive bias $\mu_l > \mu_r$). The electron transmission
is nearly unity above the onsite energy. The larger the transmission, the
larger the current and the heat generation provided that the other
parameters remain unchanged.  These two threshold behaviours may become less
obvious when the coupling with the leads get stronger. As a result of
coupling, the discrete electron and phonon density of states (DoS) extends
to a small energy region around their discrete values. The continuous phonon
DoS leads to the broadening of the first threshold behaviour, while the
continuous electron DoS is responsible for that of the second. It is
smoothed out when the coupling is large enough (Fig.~\ref{fig:ecouple}).
Only electrons whose energies are within the broadened energy spectrum can
tunnel across the central atom. The heat generation reaches maximum when the
electron states in one lead are all occupied in this energy range, while
those in the other are all empty.

The electron-lead coupling not only leads to the electron level broadening,
but it also influences the electron tunneling time. The larger this
coupling, the less time electrons spend in the central region.  In Fig.~\ref{fig:ecouple}, we show the
heat generation and the atom temperature for a single-atom structure under different electronic coupling strengths. The definition of temperature is ambiguous in
nanostructures\cite{cahill:793}. Here we use the method proposed in Ref.~\onlinecite{MGalperion07-mtj}.  We can only see
one threshold behaviour at about $0.2$ V, which is smoothed out when the
coupling is larger than $0.2$ eV. The temperature and the heat generation
show similar trends. The saturate voltage of heat generation increases with
the strengthen of the electron-lead coupling.  This is due to the coupling
induced atomic level broadening.  The decrease of the heat generation and
temperature with increasing electron-leads coupling can be easily
understood. The larger this coupling, the less time electrons spend at the
central atom. Since the electron-phonon interaction takes place there, the
heat generation decreases.  We also show the heat generation as a function
of electron-lead coupling in the inset of the lower panel. The applied
voltage is $0.3$ V. On one side, when the coupling is too small, few
electrons can tunneling through the atom. The heat generation is
small. On the other, when the coupling is very large, the electron tunneling
process is too quick for the phonons to interact with the electrons. The
heat generation is also small. It has a maximum value at some moderate
coupling strength. This is different from the electrical current, which increases monotonously with the increase of coupling strength.

\begin{figure}[!htbp]
\includegraphics[scale=1.2]{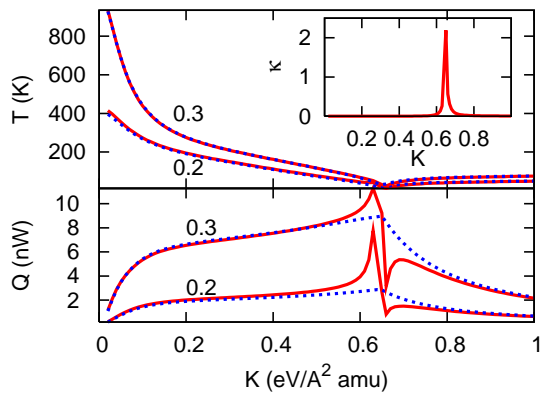}
\caption{\label{fig:phcouple}Heat generation $Q$ and the atom
temperature $T$ as a function of the atom-lead coupling $K^{CL}_{10}=K^{RC}_{21}=K$. Dashed and dotted lines correspond to results of
SCBA and BA, respectively. Other parameters are the same with Fig.~\ref{fig:scvsnosc}. The inset shows the thermal conductance $\kappa$ as a function of $K$. The unit is $1\times10^{-12}$ W/K.}
\end{figure}

The atom-lead coupling determines how well the generated heat can be
conducted into the surrounding leads. One of the important reasons why we
are interested in the heat generation in nanostructures is that it may
leads to temperature increase and even structure breakup. To study the
temperature change, we need to take into account not only the heat
generation, but also the heat conduction into the leads. In the simplest
one-atom structure, the heat conductance is mainly determined by the
atom-lead coupling.  Our model includes this intrinsically. Figure
\ref{fig:phcouple} shows the heat generation and the atom temperature as a
function of atom-lead coupling.  For the heat generation, the BA and SCBA
results show large difference around the resonant position, which
corresponds to a perfect atomic junction.  For the atom temperature, BA and SCBA
give almost the same results. In the case of a perfect junction, the
heat generation reaches its maximum value, while the atom temperature is the
lowest. The reason is that the perfect junction has the best heat conductance. When the atom-lead coupling is weak, the heat
generation is small. But the poor heat conductance can still result in a
much higher temperature than the surrounding leads. We also show the heat
conductance as a function of atom-lead coupling in the inset of the upper
panel, which shows a sharp peak at resonance. 

\begin{figure}[!htbp]
\includegraphics[scale=1.2]{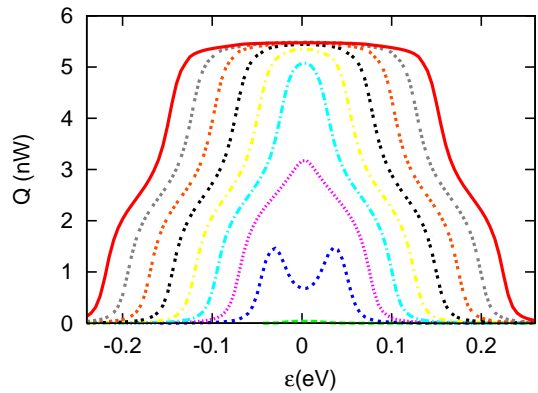}
\caption{\label{fig:eposition}Heat generation $Q$ as a function of electrical onsite energy $\varepsilon_1^C$ under different biases. From the inner to the outer side, the applied biases are $V = 0.05$, $0.10$, $0.15$, $0.20$, $0.25$, $0.30$, $0.35$, $0.40$, $0.45$, and $0.50$ V, respectively.}
\end{figure}

In Fig.~\ref{fig:eposition}, we show the heat generation as a function of
electron onsite energy at different biases. We assume that we can tune the
the onsite energy via a gate voltage. When the applied bias is less
than the phonon energy, there will be no heat generation. When the bias
energy is slightly larger than the phonon energy and less than $2\omega$,
there are two energy positions where the heat generation is the largest.
These two peaks are approximately at $-0.5 eV+\omega$ and $0.5 eV-\omega$.
They merge into a single one at a bias of $eV = 2\omega$ until it reaches
saturation. After that, this peak broadens, and becomes ladders.  All these
behaviour can also be explained by the analytical result of Eq.~(\ref{eq:heat3}). 

\begin{figure}[!htbp]
\includegraphics[scale=1.2]{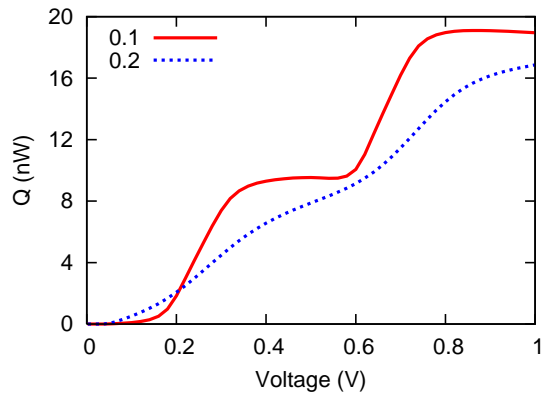}
\caption{\label{fig:twow}Heat generation $Q$ as a function of applied
voltage for a two-atom structure. The two-atom onsite energy is $\varepsilon^C = 0.2$ eV, the hopping
energy is $t^{C} = 0.1$ eV, and the spring constant is $K^{C} = 0.654$
eV/(\AA$^2\cdot$amu). The two leads are identical.
The electron onsite energy is $\varepsilon^L=\varepsilon^R = 0$, and the
hopping energy is $t^{L}=t^{R}=0.5$ eV. Their spring constants are the same
as the central region. The non-zero coupling couplings with the leads are
$K_{10}^{CL} = K_{32}^{RC} = 0.327$ eV/(\AA$^2\cdot$amu), and
$t^{CL}_{10}=t^{RC}_{32}=0.1$ (solid), $0.2$ eV (dashed), respectively.  
}
\end{figure}

In Fig.~\ref{fig:twow} we show the heat generation of a two-atom structure ($n=2$ in Fig.~\ref{fig:show}).
The central region has two identical atoms. Interaction between them leads
to two discrete energy levels. One is at $0$ eV, and the other at $0.4$ eV.
When the electrical coupling between the leads and the central region is
small ($0.1$ eV), additional to the threshold behaviour at $eV=\omega$, there
are two ladders corresponding to the phonon assisted resonant tunneling
across the two electrical levels.  If the electrical coupling gets larger
($0.2$ eV), the two ladders broaden out.  Again this is attributed to the
coupling induced level broadening. The heat generation for the two-atom
structure is much larger than that of a single-atom structure. The more the
electrical levels, the larger the electrical current and heat generation.
It is worth noting that for multi-atom structures the distribution of the
electrostatic potential may influence the results
significantly\cite{segal:3915}. In the above calculation, we assume that the
two electrical levels don't change with the applied bias, and that we can
tune their positions via a gate voltage.

\begin{figure}[!htbp]
\includegraphics[scale=1.2]{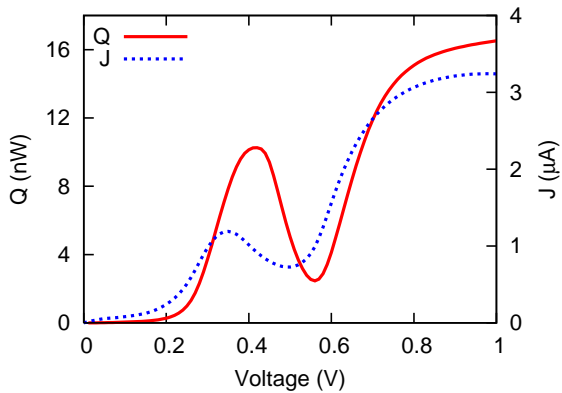}
\caption{\label{fig:semi}Heat generation $Q$ and electrical current $J$ as a function of applied voltage
for a single-atom structure. The left lead is a semiconductor. Its alternating onsite energies are $-0.2$ and $-0.1$ eV, respectively. The chemical potential is $\mu^L = 0.05$ eV higher than the conduction
band bottom, which corresponds to $n$-type doping. Other parameters are 
$K^{CL}_{10} = K^{RC}_{21} = 0.4$ eV/(\AA$^2\cdot$amu), $t^{CL}_{10}=t^{RC}_{21}=0.1$ eV, $\varepsilon^C_1 = 0.1$ eV, $K^{C}_{11}=0.654$ eV/(\AA$^2\cdot$amu), $\varepsilon^R = -0.05$ eV, $t^{L}=t^R = 0.5$ eV, and $K^L=K^R=0.654$ eV/(\AA$^2\cdot$amu).
}
\end{figure}

If one of the metallic leads is replaced by a semiconductor, there will be
some new features in the electrical current and the heat generation. In our
simple model, we can alternate the electron onsite energies between two
values to mimic a simple semiconductor (Appendix \ref{sec:se}). In Fig.~\ref{fig:semi} we show the heat generation and the electrical current for
such kind of structure. The alternating onsite energies of the left lead are
$-0.1$ and $-0.2$ eV, respectively. This produces an energy band-gap of
$0.1$ eV. Other parameters are given in the figure caption. We can see that
there appears negative differential conductivity in the current-voltage
characteristics due to the semiconductor band-gap. This qualitatively agrees
with the experimental\cite{GuisingerN.P._nl0348589} and first-principle\cite{rakshit:125305} studies. The heat generation curve
is slightly different. Additional to its threshold behaviour, the peak and
valley positions are also different. The electrical current has a peak
when the chemical potential of the lead is aligned with the central
electrical level, while the peak of the heat generation shifts to the right
by one phonon energy. This corresponds to the phonon-assisted resonant
tunneling. The current and heat generation decrease when the single
electrical level is within the band-gap of the left lead. The peak-to-valley
ratio depends on the coupling with  the semiconductor lead. In the limit of
small band-gap and large coupling, we recover the metallic lead results.

\section{Conclusion}
\label{sec:con}

We studied the coupled electron and phonon transport in 1D atomic
junctions in the weak electron-phonon interaction regime. Base on the
nonequilibrium Green's function method, we derived the electrical, energy
current of the coupled electron-phonon system, and the energy exchange between
them. We showed that the SCBA
conserves the energy current. Using this formalism, we studied the heat
generation in one- and two-atom structures coupling with different leads under
a broad range of parameters.  Especially, we studied the influence of the
thermal transport properties on the heat generation and atom temperature of the
central region. The results on semiconductor leads agree qualitatively with the
experimental and first-principle studies. This model can be easily extended to
study more realistic structures such as molecular transport junctions and
metallic nanowires. The electron, phonon Hamiltonian, their interaction and
lead-coupling matrices can all be obtained from first-principle
calculations\cite{PhysRevB.63.245407,wang:033408,frederiksen:256601}.  The
surface Green's functions for bulk leads can be computed by recursive
method\cite{PhysRevB.63.245407,wang:033408}. It is also possible to include
the electron-electron and the phonon-phonon
interactions\cite{wang:033408,PhysRevB.68.245406}.

\begin{acknowledgments}
We thank Baowen Li, Sai Kong Chin, Jian Wang, and Nan Zeng for discussions. 
This work is supported in part by a Faculty Research Grant of the National
University of Singapore.
\end{acknowledgments}
\appendix
\section{Surface Green's functions of the 1D lead}
\label{sec:se}

In this Appendix, we show that for the 1D tight-binding model the lead
self-energies can be expressed analytically\cite{wang:033408}. The electron
and phonon self-energies are similar in their form. Here we take electrons
as an example, and give the phonon results directly. We assume that the
onsite energies of the electrons alternate between $\varepsilon_1$ and
$\varepsilon_2$. The hopping energy is $t^\alpha_{ij}=t_0$. If
$\varepsilon_1 = \varepsilon_2$, we get a continuum band. This corresponds
to a metallic lead. If they are not equal, we get two bands with a band gap.
We can take the lower as the valence band (VB), and the upper as the conduction
band (CB). We use this method to mimic a semiconductor lead. In this case, the
semi-infinite lead has two electron states in each period. In the tight-binding
model, only the left- (right-) most state of the central region is coupled to the left
(right) lead.  So we only need to know the surface Green's function, e.g.,
for the left lead it is $g_0=g^r_{00}$. We assume the retarded
Green's function is
\begin{equation}
g^r_{ij} = \left\{\begin{array}{ll} c_1 \lambda^{i-j} & \quad \textrm{state 1}, \\
c_2 \lambda^{i-j} &  \quad \textrm{state 2}. \end{array}\right. 
\end{equation}
Putting it into the definition of the retarded Green's functions
$[(\varepsilon+i\eta) I - H]g^r = I$, we have
\begin{equation}
	-t_0 c_1 + (\varepsilon+i\eta -\varepsilon_2)c_2 -t_0c_1\lambda = 0,
	\label{eq:e1}
\end{equation}
\begin{equation}
	-t_0 c_2 + (\varepsilon+i\eta -\varepsilon_1)c_1\lambda -t_0c_2\lambda = 0.
	\label{eq:e2}
\end{equation}
From Eqs. (\ref{eq:e1}-\ref{eq:e2}), we get an equation for $\lambda$
\begin{equation}
	\lambda^2 + \left[2-\frac{(\varepsilon+i\eta-\varepsilon_1)(\varepsilon+i\eta-\varepsilon_2)}{t_0^2}\right]\lambda + 1 = 0.
	\label{eq:l1}
\end{equation}
The condition that Eq.~(\ref{eq:l1}) has travelling wave solutions gives the
dispersion relation
\begin{equation}
	\begin{array}{ll}
		\frac{(\varepsilon_1+\varepsilon_2)-\sqrt{(\varepsilon_1-\varepsilon_2)^2+16t_0^2}}{2} \le \varepsilon \le \varepsilon_1 &  \quad \textrm{(VB),}\\
	\varepsilon_2 \le \varepsilon \le \frac{(\varepsilon_1+\varepsilon_2)+\sqrt{(\varepsilon_1-\varepsilon_2)^2+16t_0^2}}{2} & \quad \textrm{(CB).}
	\end{array}
	\label{eq:dis1}
\end{equation}
We assume $\varepsilon_1\le \varepsilon_2$ without loss of generality. The energy band-gap
is $\varepsilon_2-\varepsilon_1$. If they are equal, the two bands
merge into one, which corresponds to a metallic lead. 

For the surface Green's function of the left lead, we also have
\begin{equation}
	(\varepsilon + i\eta -\varepsilon_1)c_1 - t_0c_2 = 1.
	\label{eq:e3}
\end{equation}
From Eqs. (\ref{eq:e1},\ref{eq:e3}), we get 
\begin{equation}
	g_0 = \left\{\begin{array}{ll}\frac{\varepsilon + i\eta -\varepsilon_2}{(1+\lambda)t_0^2} &  \quad \textrm{(VB),}\\
		\frac{\varepsilon + i\eta - \varepsilon_1}{(1+\lambda)t_0^2} & \quad \textrm{(CB).} 
	\end{array}\right.
	\label{eq:surg}
\end{equation}
$|\lambda|\ge1$ is one of the roots of Eq.~(\ref{eq:l1}). The surface Green's function of the right lead is identical. 

We can also alternate the atom masses to generate a phonon band-gap. In our model the mass change will modify the renormalized spring constants. The diagonal elements of the dynamical matrix will be two alternating values $K^\alpha_{ii} = 2k_1$ or $2k_2$, while the off-diagonal elements will be a single value $K^\alpha_{ij} = -\sqrt{k_1k_2}$, where $|i-j|=1$.
If we assume
that $k_2 \ge k_1$, the acoustic band (AB) is $0<\omega^2<2k_1$, and the
optical band (OB) $2k_2 < \omega^2 < 2(k_1+k_2)$. The surface Green's function 
is
\begin{equation}
	d_0 = \left\{\begin{array}{ll}\frac{\Omega_2 }{(1+\lambda)k_1k_2} & \quad \textrm{(AB),} \\
		\frac{\Omega_1}{(1+\lambda)k_1k_2} &  \quad \textrm{(OB),} \end{array}\right.
	\label{eq:sephb}
\end{equation}
where $\Omega_n = (\omega+i\eta)^2 - 2k_n$.  $|\lambda| \ge 1$ is one of the roots of
\begin{equation}
	\lambda^2 + \left(2-\frac{\Omega_1\Omega_2}{k_1k_2}\right)\lambda+1=0.
	\label{eq:pheq}
\end{equation}
In all the simulation results of present paper, the two spring constants
are equal ($k_1=k_2$), which correspond a single continuum phonon band. The
electron onsite energies are also equal ($\varepsilon_1=\varepsilon_2$)
except in Fig.~\ref{fig:semi}, where we set $\varepsilon_1=-0.2$ eV and
$\varepsilon_2=-0.1$ eV to mimic a semiconductor lead.

\section{Energy current conservation}
\label{sec:cons}
In this Appendix, we justify that the SCBA 
satisfies the energy current conservation. The justification of the electrical
current conservation is given in the Refs.~\cite{frederiksen-master,viljas:245415}. What we need to prove is that
\begin{eqnarray}
	\sum_\alpha (J^{\rm E,e}_\alpha+J^{\rm E,ph}_\alpha)  = 0.
	\label{eq:cont1}
\end{eqnarray}
The electron part is
\begin{equation}
	\sum_\alpha J^{\rm E,e}_\alpha = \sum_\alpha \int \frac{d\varepsilon}{2\pi} \varepsilon~\mathrm{Tr}\{G^>(\varepsilon)\Sigma_\alpha^<(\varepsilon)-G^<(\varepsilon)\Sigma_\alpha^>(\varepsilon)\}.
	\label{eq:ee1}
\end{equation}
Using the important relation\cite{frederiksen:256601,frederiksen-master}
\begin{equation}
	\label{eq:tcont}
	\mathrm{Tr}\left\{G^>\Sigma_t^<-G^<\Sigma_t^>\right\} = 0,
\end{equation}
we get 
\begin{equation}
	\sum_\alpha J^{\rm E,e}_\alpha = - \int \frac{d\varepsilon}{2\pi} \varepsilon~\mathrm{Tr}\{G^>(\varepsilon)\Sigma_{\rm eph}^<(\varepsilon)-G^<(\varepsilon)\Sigma_{\rm eph}^>(\varepsilon)\}.
	\label{eq:ee2}
\end{equation}
The Hartree term does not contribute to the current directly. It's just like
a static potential which only modifies the Green's function.  Putting the Fock self-energy into Eq.~(\ref{eq:ee2}), we have
\begin{eqnarray}
	\label{eq:selcontfock}
	-Q&=&\sum_\alpha J^{E,e}_\alpha \nonumber \\
	&=&-i\int \frac{d\varepsilon}{2\pi} \int \frac{d\omega}{2\pi} \varepsilon \left [G^>_{nm}(\varepsilon)M_{mi}^kD_{kl}^<(\omega)G^<_{ij}(\varepsilon-\omega)M_{jn}^l\right. \nonumber \\
	&&\left. -G^<_{nm}(\varepsilon)M_{mi}^kD_{kl}^>(\omega)G^>_{ij}(\varepsilon-\omega)M_{jn}^l\right]. 
\end{eqnarray}
Sum over all the indices is assumed. The heat generation $Q$ is the energy
decrease of the electron system, which should also be the energy increase of
the phonon system.  Replacing $\omega$ by $-\omega$, using the symmetric
properties of the phonon Green's functions\cite{wang:033408}, replacing
$\varepsilon$ by $\varepsilon-\omega$, and finally changing dummy variables,
we get 
\begin{eqnarray}
	\label{eq:selcontfock2}
	&&i\int \frac{d\varepsilon}{2\pi} \int \frac{d\omega}{2\pi} \varepsilon \left[G^<_{nm}(\varepsilon)M_{mi}^kD_{kl}^>(\omega)G^>_{ij}(\varepsilon-\omega)M_{jn}^l\right] \nonumber \\
	 &=&i\int \frac{d\varepsilon}{2\pi} \int \frac{d\omega}{2\pi} \varepsilon \left [G^<_{nm}(\varepsilon)M_{mi}^kD_{kl}^>(-\omega)G^>_{ij}(\varepsilon+\omega)M_{jn}^l\right] \nonumber \\
	&=&i\int \frac{d\varepsilon}{2\pi} \int \frac{d\omega}{2\pi} (\varepsilon-\omega) \left [G^<_{nm}(\varepsilon-\omega)M_{mi}^kD_{lk}^<(\omega)G^>_{ij}(\varepsilon)M_{jn}^l\right] \nonumber \\
	&=&i\int \frac{d\varepsilon}{2\pi} \int \frac{d\omega}{2\pi} (\varepsilon-\omega) \left [G^>_{nm}(\varepsilon)M_{mi}^kD_{kl}^<(\omega)G^<_{ij}(\varepsilon-\omega)M_{jn}^l\right].\nonumber\\
        &&
\end{eqnarray}
Putting Eq.~(\ref{eq:selcontfock2}) back into Eq.~(\ref{eq:selcontfock}), we get 
\begin{eqnarray}
	\label{eq:selconte}
	 -Q&=&\sum_\alpha J_\alpha^{\rm E,e} \nonumber \\
	 &=&-i\int \frac{d\varepsilon}{2\pi} \int \frac{d\omega}{2\pi} \omega \left [G^>_{nm}(\varepsilon)M_{mi}^kD_{kl}^<(\omega)G^<_{ij}(\varepsilon-\omega)M_{jn}^l\right] \nonumber \\
	 &\ne&0.
\end{eqnarray}
For the phonon energy current we have
\begin{eqnarray}
	\label{eq:sephonon}
	Q&=&\sum_\alpha J^{\rm E,ph}_\alpha \nonumber \\
	&=&i\int \frac{d\varepsilon}{2\pi} \int \frac{d\omega}{4\pi} \omega \left [D^>_{nm}(\omega)M_{lk}^mG_{ki}^<(\varepsilon)G^>_{jl}(\varepsilon-\omega)M_{ij}^n \right. \nonumber \\
	&& \left. -D^<_{nm}(\omega)M_{lk}^mG_{ki}^>(\varepsilon)G^<_{jl}(\varepsilon-\omega)M_{ij}^n\right]. 
\end{eqnarray}
Following the same procedure as electrons, finally we get 
\begin{eqnarray}
	\label{eq:selconte2}
	Q && = \sum_\alpha J_\alpha^{\rm E,ph} \nonumber \\
	&& = i\int \frac{d\varepsilon}{2\pi} \int \frac{d\omega}{2\pi} \omega \left [G^>_{nm}(\varepsilon)M_{mi}^kD_{kl}^<(\omega)G^<_{ij}(\varepsilon-\omega)M_{jn}^l\right] \nonumber \\
	&& \ne 0.
\end{eqnarray}
So we still have 
\begin{equation}
	\label{eq:selconte3}
	\sum_\alpha \left(J_\alpha^{\rm E,e}+J_\alpha^{\rm E,ph}\right) = 0.
\end{equation}
Eqs. (\ref{eq:selconte}, \ref{eq:selconte2}) give the energy exchange between
the electron and the phonon system, which is also the heat generation of the atomic
junction. Replacing $D^<$, $G^<$ by $D_0^<$, $G_0^<$ in Eq.~(\ref{eq:selconte}), and $G^{>}$, $G^{<}$ by $G_0^>$, $G_0^<$ in Eq.~(\ref{eq:selconte2}), we  get the results under BA. We can find that the
energy increase of the phonons does not equal to the energy decrease of the
electrons under BA.

\section{Analytical result at zero temperature}
\label{sec:ana}
At zero temperature, we can get an analytical expression for heat generation
in a single-atom structure by using the bare Green's functions in Eq.~(\ref{eq:selconte}, \ref{eq:selconte2}). We only take into account the
imaginary part of the lead self-energies and ignore their energy dependence
(the wide-band limit)\cite{PhysRevLett.68.2512}. Finally, we assume that the
phonons are in their equilibrium states. Under these approximations, the
heat generation is (assuming $eV \ge \omega_0$) 
\begin{eqnarray}
	Q&\approx&\frac{1}{2} M^2 \Gamma_L\Gamma_R \int_{-\frac{eV}{2}+\omega_0}^{\frac{eV}{2}} \frac{d\varepsilon}{2\pi} \nonumber \\
	&& \times\frac{1}{\left[ (\varepsilon-\varepsilon_0)^2 +\Gamma^2/4\right]\left[ (\varepsilon-\varepsilon_0-\omega_0)^2 +\Gamma^2/4\right]}\nonumber \\
	&=&
	\frac{M^2\Gamma_L\Gamma_R}{4\pi(\omega_0^2+\Gamma^2)}
	\left\{\frac{1}{\omega_0}\left[\mathrm{ln}\left(\frac{(eV/2-\varepsilon_0)^2+\Gamma^2/4}{(eV/2-\varepsilon_0-\omega_0)^2+\Gamma^2/4}\right)\right.\right.\nonumber \\
	&& \left.\left.-\mathrm{ln}\left(\frac{(-eV/2-\varepsilon_0+\omega_0)^2+\Gamma^2/4}{(-eV/2-\varepsilon_0)^2+\Gamma^2/4}\right)\right]\right. \nonumber\\
	 && \left.+\frac{2}{\Gamma}\left[\mathrm{arctan}\left(\frac{eV/2-\varepsilon_0}{\Gamma/2}\right)+\mathrm{arctan}\left(\frac{eV/2-\varepsilon_0-\omega_0}{\Gamma/2}\right)\right.\right.\nonumber\\
&& \left.\left.-\mathrm{arctan}\left(\frac{-eV/2-\varepsilon_0}{\Gamma/2}\right)\right.\right.\nonumber\\
	 && \left.\left.-\mathrm{arctan}\left(\frac{-eV/2-\varepsilon_0+\omega_0}{\Gamma/2}\right)\right]\right\}.
	\label{eq:heat3}
\end{eqnarray}
$\omega_0$ is the phonon energy, $\varepsilon_0$
is the electron onsite energy, $V$ is the applied bias, and $\Gamma =
\Gamma_L+\Gamma_R$. The heat generation is zero when $eV \le \omega_0$. Equation~(\ref{eq:heat3}) can reproduce most the qualitative features of heat
generation in a single atom, except that it does not take into account heat
conduction in the phonon system. 

%
\bibliography{nus}

\begin{thebibliography}{42}
\bibitem[{not({\natexlab{b}})}]{note2}
\bibinfo{note}{We denote electron energy as $\varepsilon$, and phonon energy as
  $\omega$, except in the unified expressions including both electrons and
  phonons, i.e., Eq. (22).}
\end{thebibliography}
\end{document}